\documentclass[12pt,a4]{article}
\usepackage{a4wide}
\usepackage{latexsym}
\usepackage{epsf}
\usepackage{amssymb}
\usepackage{cite}

\usepackage{amsmath}
\usepackage{slashed,epsfig}

\makeatletter \@addtoreset{equation}{section} \makeatother

\newsavebox{\uuunit}
\sbox{\uuunit}
    {\setlength{\unitlength}{0.825em}
     \begin{picture}(0.6,0.7)
        \thinlines
        \put(0,0){\line(1,0){0.5}}
        \put(0.15,0){\line(0,1){0.7}}
        \put(0.35,0){\line(0,1){0.8}}
       \multiput(0.3,0.8)(-0.04,-0.02){12}{\rule{0.5pt}{0.5pt}}
     \end {picture}}

\begin{document}

\begin{flushright}
\date \\
\normalsize
\end{flushright}

\begin{center}


\vspace{1.7cm}

{\Large {\bf Casimir effect in a six-dimensional vortex scenario}}

\vspace{1.5cm}


{\large Rom\'an Linares, Hugo A. Morales-T\'ecotl\footnote{Associate
member of the Abdus Salam ICTP, Trieste Italy.} and Omar Pedraza}

\vskip 1truecm

\small
{\it Departamento de F\'{\i}sica, Universidad Aut\'onoma Metropolitana Iztapalapa, \\
San Rafael Atlixco 186, CP 09340, M\'exico D.F., M\'exico.}

E-mail: {\tt lirr,hugo,omp@xanum.uam.mx} \vskip 0.2cm

\vspace{.7cm}


{\bf Abstract}

\end{center}

\begin{quotation}

\small Recently Randjbar-Daemi and Shaposhnikov put forward a
4-dimensional effective QED coming from a Nielsen-Olesen vortex
solution of the abelian Higgs model with fermions coupled to gravity
in $D=6$. However, exploring possible physical consequences of such
an effective QED was left open. In this letter we study the
corresponding effective Casimir effect. We find that the extra
dimensions yield fifth and third inverse powers in the separation
between plates for the modified Casimir force which are in conflict
with known experiments, thus reducing the phenomenological viability
of the model.
\end{quotation}

\newpage

\pagestyle{plain}

\section{Introduction}

The idea that our observable 4-dimensional universe may be a brane
extended in some higher-dimensional space-time has been attracting
interest for many years
\cite{Akama:1982jy,Rubakov:1983bb,Visser:1985qm}. Roughly speaking,
there exist two different approaches to implement this idea. One
approach is to start with theories that incorporate gravity in a
reliable manner such as string theory/M-theory
\cite{Horava:1996ma,Horava:1995qa}. Almost all the known examples of
these kind of theories are naturally and consistently formulated in
higher-dimensions. For instance, it is possible to include chiral
fermions by considering intersecting $D$-branes
\cite{Cremades:2002dh,Kokorelis:2002qi,Blumenhagen:2005mu}. The
second approach follows more phenomenological lines and is often
based on simplified field-theoretical models which have recently led
to new insights on whether they may help to solve long-standing
problems of particle theory such as the hierarchy problem, the
cosmological constant problem, etc.
\cite{Arkani-Hamed:1998rs,Antoniadis:1998ig,Arkani-Hamed:1998nn,Randall:1999ee,Randall:1999vf,Brevik:2000vt,Barrow:1990nv,Elizalde:2004mq}
(see for instance the comprehensive reviews
\cite{Perez-Lorenzana:2005iv,Feruglio:2004zf}).

An important problem in the field theory approach is to find natural
mechanisms for localization of the different fields to 4-dimensional
space-time. There exist many models that can achieve the
localization of scalar and fermionic fields, however, the
localization of gauge fields is not an easy challenge to tackle
\cite{Gherghetta:2000qi,Dubovsky:2000av}. Recently, starting from a
Higgs model with fermions coupled to gravity in $D=6$,
Randjbar-Daemi and Shaposhnikov \cite{Randjbar-Daemi:2003qd}
constructed an effective quantum electrodynamics in 4-dimensional
space-time, with fermionic and gauge functions spread on the
transverse direction in a small region in the vicinity of the core
of a Nielsen-Olesen vortex. This construction is possible because
the vortex solution
\cite{Giovannini:2002sb,Giovannini:2001hh,Randjbar-Daemi:2002pq},
admits gravity localization \cite{Giovannini:2001hh} and contains
the massless $U(1)$ gauge field, which is a mixture of a graviton
fluctuation and the original $U(1)$ gauge field fluctuation forming
the Nielsen-Olesen vortex.

Since the 4-dimensional effective QED owns many non trivial
properties, despite all the theoretical interest it is natural to
ask ourselves how far we can go with this model and compute its
consequences in low/high energy physics. In doing this there exists
the additional possibility of saying something about the potential
detectability of extra dimensions by measuring effects which for
this particular model have not been discussed to the best of our
knowledge. The aim of this letter is to analyze, the Casimir effect
between parallel plates in the context of the effective QED of
 \cite{Randjbar-Daemi:2003qd}.

The standard Casimir effect between parallel, uncharged, perfectly
conducting plates is understood on the basis of the ordinary
4-dimensional QED. For flat plates separated by a distance $l$, the
force per unit area $A$ is given by
$F(l)/A=-\frac{\pi^2}{240}\frac{\hbar c}{l^4}$. This relationship is
derived considering the electromagnetic mode structure between the
two parallel plates, as compared to the mode structure when the
plates are infinitely far apart, and by assigning a zero-point
energy of $\hbar \omega/2$ to each electromagnetic mode (photon)
\cite{Casimir:1948dh}. The change in the total energy density
between the plates, as compared to the free space, as a function of
the separation $l$, leads to the force of attraction. The only
fundamental constants that enter into the expression of the force
are $\hbar$ and $c$. The electron charge $e$ is absent, implying
that the electromagnetic field is not coupled to matter. The role of
$c$ is to convert the electromagnetic mode wavelength, as determined
by $l$, to a frequency, while $\hbar$ converts the frequency to an
energy. The Casimir effect has also been obtained for other fields
and other geometries of the bounding surfaces which may be described
by real material media, with electromagnetic properties
\cite{Milton:2001yy}.

The Casimir effect, on the other hand, has received great deal of
attention within theories and models with extra dimensions. For
example, it has been discussed in the context of string theory
\cite{Fabinger:2000jd,Gies:2003cv,Brevik:2000fs,Hadasz:1999tr}. In
the Randall-Sundrum model, the Casimir effect has been considered to
stabilize the radion
\cite{Elizalde:2002dd,Garriga:2002vf,Pujolas:2001um,Flachi:2001pq,Goldberger:2000dv}
as well as within the inflationary brane world universe models
\cite{Nojiri:2000bz}. More recently the effect was analyzed in the
presence of compactified universal extra dimensions
\cite{Poppenhaeger:2003es}. In all these cases the boundaries in the
extra dimensions are associated to the topology of space.

In general the Casimir effect may be defined as the stress on the
bounding surface when a quantum field is confined to a finite volume
of space. In any case, the boundaries restrict the modes of the
quantum field giving rise to a force which can be either attractive
or repulsive, depending on the model, the field and the space-time
dimension.

In this letter we start with the 4-dimensional effective QED of
Randjbar-Daemi and Shaposhnikov (section 2) in order to determine
the dispersion relations of the electromagnetic modes (section 3).
This is done near the core of the vortex scenario that is meant to
represent our world. Next we proceed with the standard approach of
analysis of the Casimir effect. Namely we add up the electromagnetic
mode contributions to the energy between two parallel conducting
plates (section 4). Finally we discuss our results in section 5.

\section{$D=4$ Effective QED}\label{QED}

The starting point in our analysis is the 4-dimensional effective
QED of Randjbar-Daemi and Shaposhnikov \cite{Randjbar-Daemi:2003qd}.
This theory emerges from considering a Nielsen-Olesen vortex-type
solution of the  abelian Higgs model with fermions coupled to
gravity in $D=6$. The various field configurations of the solution
are \cite{Giovannini:2001hh}
\begin{eqnarray}
ds^2&=&\mbox{e}^{A(r)}\eta_{\mu\nu}dx^\mu dx^\nu
+dr^2+\mbox{e}^{B(r)} a^2d\theta^2 \, ,\nonumber \\
\Phi&=&f(r)\mbox{e}^{in\theta},
\hspace{0.5cm}aeA_{\theta}=(P(r)-n)d\theta \, ,
\end{eqnarray}
where $\eta_{\mu \nu}$ is the $D=4$ flat metric, $e$ is the
6-dimensional gauge coupling and $a$ is the radius of $S^1$ covered
by the $\theta$ coordinate. To avoid confusion below notice the
difference between $e$ and $\mbox{e}$. The boundary conditions that
$f(r)$ and $P(r)$ must satisfy are
\begin{equation}
f(0)=0, \hspace{0.5cm} f(\infty)=f_0, \hspace{0.5cm} P(0)=n
\hspace{0.5cm} \mbox{and} \hspace{0.5cm} P(\infty)=0.
\end{equation}
On the other hand, there are solutions with different boundary
conditions for the metrical functions $A(r)$ and $B(r)$
\cite{Gherghetta:2000qi}. Among all of them, the one that localize
fields of spin 0, 1/2 and 1, near the core of the vortex satisfy the
boundary conditions
\begin{eqnarray}
A(0)&=&1, \hspace{1cm}B(r\rightarrow 0)=2\ln
\frac{r}{a}, \nonumber\\
A(r\rightarrow \infty)&=&B(r\rightarrow \infty)=-2cr,
\hspace{0.8cm}c>0,
\end{eqnarray}
where the parameters $a$ and $c$ are combinations of the
6-dimensional gravitational constant $\kappa$, the cosmological
constant and of the parameters of the abelian Higgs model
\cite{Gherghetta:2000qi}. In this case as $r\rightarrow0$ the flat
space geometry is recovered whereas for $r\rightarrow \infty$ the
metric becomes AdS.

The effective QED action in this background results from a specific
mixture of the fluctuation of the 6-dimensional vector potential and
the $\theta\mu$ component of the metric
\cite{Randjbar-Daemi:2002pq}. Its explicit form is
\begin{equation}
S(W)=-\frac{\pi}{ae^2}\int_{0}^{\infty} {\mathrm d}rF(r) \int
{\mathrm d}^4x \left[(\partial_{\mu}W_{\nu})^2
+\mbox{e}^{-A}\partial_{r}W_{\mu}\partial_{r}W^{\mu}\right],
\end{equation}
where
\begin{equation}
F(r)=\mbox{e}^{\frac{B(r)}{2}}\left(P^2(r)+\frac{a^2e^2}{\kappa^2}\mbox{e}^{B(r)}\right).
\end{equation}
Notice that the quotient $\ell \equiv \kappa/e$ has dimension of
length. The equations of motion for the gauge fields are
\begin{equation}\label{eq:W}
F\partial_{\mu}\partial^{\mu}W^{\nu}+\partial_{r}\left(F\mbox{e}^{-A}\partial_{r}W^{\nu}\right)=0.
\end{equation}

\section{Dispersion relations}\label{}

As discussed in \cite{Randjbar-Daemi:2002pq}, a 4-dimensional
effective low energy theory can arise if two conditions are
satisfied:

(i) The spectrum contains normalizable zero (or small mass) modes of
graviton, gauge, scalar and fermion fields, with wave functions of
the type $\mbox{e}^{ip_\mu x^\mu}\psi(y^m)$; $y^m$ here represent
the extra dimensional coordinates.

(ii) The effects of higher modes should be experimentally
unobservable at low enough energies, i.e. there should be a mass gap
between the zero modes and excited states. Another possibility is
that extra, unwanted modes may be light but interact very weakly
with the zero modes.

In this section we shall analyze the implications of the first
condition for the zero modes of the gauge fields i.e. we shall
consider a wave function of the form
$W^{\mu}(x,r)=R^{\mu}(r)\,\mbox{e}^{ip_\nu x^\nu}$, where $p_\mu$ is
the 4-dimensional wave vector. With this form of the wave function
and proposing the change of variable $R=H(F\mbox{e}^{-A})^{-1/2}$,
the equation of motion for each component of $W$, Eq. (\ref{eq:W}),
is rewritten as a Schr\"odinger like equation
\begin{equation}\label{Schro}
\left(-\frac{1}{2}\frac{d^2}{dr^2}+V(r)\right)H(r)=-\frac{p^2\,\mbox{e}^A}{2}H(r),
\end{equation}
where the potential is given by
\begin{equation}
V(r)=\frac{1}{4}\frac{d^2}{dr^2}\ln (F\mbox{e}^{-A})+\frac{1}{8}
\left( \frac{d}{dr}\ln ( F\mbox{e}^{-A})\right)^2.
\end{equation}
Because the localization of the gauge fields is given near the core
of the vortex, we are interested in solutions where $r\rightarrow
0$. In this limit the potential is
\begin{equation}
V(r\rightarrow 0) \approx - \frac{1}{8r^2}+\frac{1}{n^2\ell^2}.
\end{equation}
and the equation (\ref{Schro}) becomes
\begin{equation}\label{eqtosolve}
\left(\frac{d^2}{dr^2}+\frac{1}{4r^2}+ k_r^2 \right)H(r)=0,
\hspace{0.5cm} \mbox{where}, \hspace{0.5cm}
k_r^2=-\left(p^2\mbox{e}+\frac{2}{n^2\ell^2}\right).
\end{equation}
Hence, in order to find the explicit functional form of the gauge
fields near the core of the vortex it is necessary to solve the
1-dimensional radial Schr\"odinger equation with an attractive
potential proportional to $1/r^2$. The solutions to the equation
(\ref{eqtosolve}) have been already discussed \cite{Case:1950an} and
it has been shown that their properties depend strongly on the value
$\lambda$ --the coefficient of the term $1/r^2$. When $\lambda <
1/4$ and the boundary condition is integrability -not finiteness-
all negative energies are allowed \cite{Morse}. For $\lambda > 1/4$,
the requirement that the state functions for bound states be a
mutually orthogonal set imposes a quantization of energy which does
not uniquely fix the levels but the levels relative to each other
\cite{Case:1950an}.

It is remarkable that the value of $\lambda$ in (\ref{eqtosolve})
takes the critical value $1/4$. The solution to the differential
equation in this case is given by
\begin{equation}
H=c_1rh^{(1)}_{-\frac{1}{2}}\left(irk_r
\right)+c_2rh^{(2)}_{-\frac{1}{2}}\left(irk_r  \right),
\end{equation}
where $h^{1}$ and $h^{(2)}$ are the spherical Hankel functions.
These functions behave in the limit $r\rightarrow0$ as
\begin{equation}
h^{(1)}_{-\frac{1}{2}}(r)\approx\sqrt{\frac{\pi}{2r}}\,\frac{2i}{\pi}\ln
r\qquad, \hspace{1cm} h^{(2)}_{-\frac{1}{2}}(r)\approx
-\sqrt{\frac{\pi}{2r}}\,\frac{2i}{\pi}\ln r.
\end{equation}
Thus if the boundary condition is integrability \cite{Morse}
\begin{equation}
\int dr R^2 \sim\int r\left(\ln rk_r\right)^2< \infty ,
\end{equation}
all negative energies are allowed and therefore the dispersion
relation is
\begin{equation}\label{dispersion}
\frac{\omega^2}{c^2}=\vec{k}\,{}^2+\frac{2}{\mbox{e}}\,k_r^2+\frac{2}{\mbox{e}}\frac{1}{n^2\ell^2}.
\end{equation}
where $\vec{k}$ is the 3-dimensional wave vector. Notice that the
dependence of the extra dimensions in the dispersion relations comes
into two different ways. There is a continuum contribution $k_r$
that comes from the radial extra dimension and there is one discrete
contribution that goes like $n^{-2}$ coming from the vortex number.
This last contribution is of a different type to the one that
emerges from a Kaluza-Klein compact extra dimension which goes like
$n^2$ \cite{Poppenhaeger:2003es}.

\section{The Casimir effect}\label{conclusions}

Once we have computed the dispersion relations we evaluate the
Casimir force between two parallel plates in the $D=4$ effective
QED. Because of the presence of the plates, we impose the standard
Dirichlet boundary condition on the wave vector in the direction
restricted by the plates: $k_N=\pi N/l$, where $l$ is the distance
between the plates. The Casimir energy between the plates is
obtained by summing up the zero-point energy per unit area, where
the frequency of the vacuum fluctuations is, according to
(\ref{dispersion}),
\begin{equation}
{\omega}_{k_\perp,\, k_r,\, N,
n}=c\sqrt{k_{\perp}^2+\frac{2}{\mbox{e}}\,k_r^2+\frac{\pi^2{N}^2}{l^2}+\frac{2}{\mbox{e}}
\frac{1}{n^2 \ell^2}},
\end{equation}
with $k_\perp = \sqrt{k_1^2+k_2^2}$. $k_1$ and $k_2$ are the wave
vector components in the direction of the unbounded space
coordinates along the plates. Each of these modes contributes an
energy $\hbar\, \omega/2$. Therefore the energy between plates reads
\begin{equation}\label{eplates}
E_{plates}=\frac{\hbar \, L^2 ap}{2} \int\frac{d^2k_\perp
dk_r}{(2\pi)^3} \sum_{n=1,N=1}^{\infty} \omega_{k_\perp,\, k_r,\, N,
n},
\end{equation}
where $L^2$ is the area of the plates and the parameter $a$,
measuring the  size of the vortex'es core, appears associated to the
integration in $k_r$. The factor $p$ indicates the possible
polarization of the photon. In our case $p=4$.

There are several ways to extract a finite value from the above
divergent sum. We shall use the one that invokes dimensional
regularization. To do so we let the transverse dimension be $d$,
which we will subsequently treat as a continuous complex variable
following \cite{Milton:2001yy}. Let us start with the expression
\begin{equation}\label{I1d}
I_1(d)= \frac{1}{2}\int\frac{d^dk}{(2\pi)^d} \sum_{n=1,N=1}^{\infty}
\sqrt{k_d^2+\frac{\pi^2{N}^2}{l^2}+\frac{2}{\mbox{e}} \frac{1}{n^2
\ell^2}}.
\end{equation}
which becomes in the limit $d\rightarrow 3$ the one in
(\ref{eplates}), namely
\begin{equation}
I_1(d=3)\equiv \frac{1}{2}\int\frac{d^2k_\perp dk_r}{(2\pi)^3}
\sum_{n=1,N=1}^{\infty} \sqrt{k_{\perp}^2+
k_r^2+\frac{\pi^2{N}^2}{l^2}+\frac{2}{\mbox{e}} \frac{1}{n^2
\ell^2}}.
\end{equation}
Using the Euler representation for the Gamma function
\begin{equation}\label{gamma}
\Gamma(z)=g^z\int_0^{\infty} \mbox{e}^{-gt}\,t^{z-1}\,dt ,
\end{equation}
the integral (\ref{I1d}) can be rewritten employing the Schwinger
proper-time representation for the square root as
\begin{equation}
I_1(d)=
\frac{1}{2}\frac{1}{\Gamma(-\frac{1}{2})}\sum_{n=1,N=1}\int\frac{d^dk}{(2\pi)^d}\int_0^\infty
\frac{dt}{t}t^{-1/2}\mbox{e}^{-t\left(k_d^2+\frac{\pi^2
N^2}{l^2}+\frac{2}{\mbox{e}}\frac{1}{\ell^2n^2}\right)}.
\end{equation}
Performing the Gaussian integral first and using (\ref{gamma}) again
we have
\begin{equation}\label{sumI1d}
I_1(d)= \frac{1}{2}\frac{1}{\Gamma(-\frac{1}{2})}
\frac{1}{(4\pi)^{d/2}} \Gamma \left(-\frac{d+1}{2}\right)
\sum_{n=1}^{\infty}\sum_{N=1}^{\infty}\left( \frac{\pi^2
N^2}{l^2}+\frac{2}{\mbox{e}}\frac{1}{\ell^2n^2}\right)^{(d+1)/2}.
\end{equation}
The double sum in (\ref{sumI1d}) is better handled by factorizing
$\left(\frac{\pi}{l}\right)^{d+1}$ and using the Epstein function
\cite{AguiarPinto:2003gq,Ambjorn:1981xw}
\begin{equation}
E_1^{M^2}(z,1)=\sum_{N=1}^{\infty}\frac{}{}\left(
N^2+M^2\right)^{-z},\qquad
M^2=\frac{2}{\mbox{e}}\frac{l^2}{\pi^2\ell^2 n^2},\qquad
z=\frac{d+1}{2} .
\end{equation}
This expression is not well defined for $\Re(z)>1/2$, however, it
can be analytically continued into a meromorphic function in the
whole complex plane, namely
\begin{equation}
E_1^{M^2}(z,1)=\frac{1}{2M^{2z}}+{\pi}^{\frac{1}{2}}
\frac{1}{2M^{2z-1}\Gamma(z)}\left[\Gamma\left(z-\frac{1}{2}\right)+4\sum_{m=1}^{\infty}\frac{1}{\left(\pi
Mm\right)^{\frac{1}{2}-z}}K_{\frac{1}{2}-z}\left({2\pi Mm}\right)
\right],
\end{equation}
where $K_{\nu}(z)$ is the modified Bessel function of second type.
Thus using the Epstein function in (\ref{sumI1d}) leads to
\begin{eqnarray}\label{eq:IK}
I_1(d)&=&\frac{1}{2}\frac{1}{\Gamma\left(-\frac{1}{2}\right)}\frac{1}{\left(4\pi\right)^{\frac{d}{2}}}
\left[\frac{1}{2}\Gamma\left(-\frac{d+1}{2}\right)
\left(\frac{2}{\mbox{e}}\frac{1}{\ell^2}\right)^{\frac{d+1}{2}}\zeta(d+1)\right.
\nonumber\\
&&\mbox{}
+\frac{l}{2\sqrt{\pi}}\Gamma\left(-\frac{d+2}{2}\right)\left({\frac{2}{\mbox{e}}}
\frac{1}{\ell^2}\right)^{\frac{d+2}{2}}\zeta(d+2)
\nonumber\\
&&\left.+
\frac{2}{\sqrt{\pi}}\left(\sqrt{\frac{2}{\mbox{e}}}\frac{1}{\ell}\right)^{\frac{d+2}{2}}\frac{1}{l^{\frac{d}{2}}}
\sum_{n,m=1}^{\infty}\left(\frac{1}{mn}\right)^{\frac{d+2}{2}}
K_{\frac{d+2}{2}}\left(\sqrt{\frac{2}{\mbox{e}}}\frac{2l}{\ell}\frac{m}{n}\right)\right].
\end{eqnarray}
Notice the first term in square brackets in (\ref{eq:IK}) is
independent of $l$ and hence it can be interpreted as a constant
energy shift upon substitution in (\ref{eplates})
\cite{Ambjorn:1981xw}. Obviously it will neither yield any
contribution to the Casimir force. Hence from now on it will be
discarded. The energy between plates takes the explicit form
\begin{equation}\label{Eneplates}
E_{plates}=\hbar\,cL^2ap\sqrt{\frac{\mbox{e}}{2}}I_1'(d\rightarrow
3),
\end{equation}
with the prime meaning we have dropped the first term in brackets in
$I_1$, Eq. (\ref{eq:IK}). Next  we compute the vacuum energy without
the plates. Appealing again to (\ref{I1d}) we have that such vacuum
energy
\begin{equation}
 E_0=\hbar\,cL^2a l\sqrt{\frac{\mbox{e}}{2}}p\,\frac{1}{2}\int\frac{d^4k}{(2\pi)^4} \sum_{n=1}^{\infty}
\sqrt{k_{\perp}^2+k_{r}^2+k_{z}^2+\frac{2}{\mbox{e}} \frac{1}{n^2
\ell^2}}
\end{equation}
becomes
\begin{equation}\label{eq:E0d4}
 E_0=\left[\hbar\,cL^2a l\sqrt{\frac{\mbox{e}}{2}}p\,\frac{1}{2}\frac{1}{\Gamma\left(-\frac{1}{2}\right)(4\pi)^{\frac{d}{2}}}\left(\frac{2}{\mbox{e}\ell^2}\right)^{(d+1)/2}
\Gamma\left(-\frac{d+1}{2}\right)\zeta(d+1)\right]_{d\rightarrow 4}
 \end{equation}
Finally, the exact Casimir energy per unit area in the vortex
scenario reads
\begin{eqnarray}\label{eq:evortex}
{\cal E}_{vortex}&=&\frac{E_{plates}-E_0}{L^2}={\cal
E}_{Casimir}\quad f\left(l,a,\ell\right)\\
{\cal E}_{Casimir}&=&-\frac{\hbar\,c\pi^2}{720l^3},\nonumber\\
f\left(l,a,\ell\right)&=& \frac{45p\left(\sqrt{\frac{2}{
\mbox{e}}}\right)^{3}}{\pi^{\frac{7}{2}}\Gamma\left(-\frac{1}{2}\right)}
\left(\frac{a}{\ell}\right)\left(\frac{l}{\ell}\right)^{\frac{3}{2}}\sum_{n,m=1}^{\infty}\left(\frac{1}{mn}\right)^{\frac{5}{2}}
K_{\frac{5}{2}}\left(\sqrt{\frac{2}{\mbox{e}}}\frac{2l}{\ell}\frac{m}{n}\right).\nonumber
\end{eqnarray}
Here we recognize ${\cal E}_{Casimir}$ as the standard
four-dimensional Casimir energy between parallel plates. Moreover,
the correction within the vortex scenario is encoded in the function
$f\left(l,a,\ell\right)$ in the form of a factor rather than an
additive term. We should also stress that to arrive to
Eq.(\ref{eq:evortex}) there occurs a cancelation between the  second
term in (\ref{eq:IK}) with the vacuum contribution with no plates,
Eq. (\ref{eq:E0d4}). To compare more neatly ${\cal E}_{vortex}$ with
the standard result it is easier to get an approximate form of it
for $l/\ell \ll1$ in the argument of the modified Bessel function
$K_{5/2}$ \cite{Morse}. This produces, to leading order in $l/\ell$,
\begin{eqnarray}
{\cal E}_{vortex}&\approx&-\alpha\frac{a}{l^4}+\beta
\frac{a}{\ell^2l^2}\\
\alpha&=&\frac{3\hbar
c\zeta(5)}{2(4\pi)^{7/2}}\Gamma\left(\frac{5}{2}\right)\sqrt{\frac{2}{\mbox{e}}},\quad
\beta=\frac{\hbar
c\zeta(2)}{32\pi^{7/2}}\zeta\left(\frac{3}{2}\right)\Gamma\left(\frac{5}{2}\right)\left(\frac{2}{\mbox{e}}\right)^{3/2}
. \nonumber
\end{eqnarray}

As for the Casimir force we obtain then
\begin{eqnarray}\label{fvortex}
F_{vortex}&=&-\frac{\partial{\cal E}_{vortex}}{\partial
l}\approx-4\alpha\frac{a}{l^5}+2\beta \frac{a}{\ell^2l^3} .
\label{eq:FCas}
\end{eqnarray}
Experimentally the Casimir force is difficult to measure because
parallelism  can not be obtained easily so it is preferable to
replace one of the plates by a metal sphere of radius $R$ where
$R>\!>l$. For such geometry the Casimir force is modified to
\begin{equation}\label{eq:fsphere}
F_{sphere}=2\pi RL^2{\cal E}_{Casimir}.
\end{equation}
The force between a  metallic sphere of diameter 196 $\mu$m and a
flat plate is measured using an atomic force microscope for
separations $l$ ranging from 0.1 to 0.9 $\mu$m
\cite{Mohideen:1998iz}. In this case the experimental uncertainty
for the Casimir force is 1.6 pN. Due to the factor correction in Eq.
(\ref{eq:evortex}) giving rise to an inverse power in the separation
between plates of $l^{-5}$ in the effective Casimir force, Eq.
(\ref{fvortex}), it is not possible to reconcile it with the
experimental results even within the error bar
\cite{Mohideen:1998iz, Decca:2005yk,Klimchitskaya:2005df} (See
\cite{Lamoreaux:2005gf} for a review of the current experimental
situation).

\section{Discussion}\label{conclusions}
In this letter we have obtained the Casimir effect corresponding to
the effective QED of Randjbar-Daemi and Shaposhnikov
\cite{Randjbar-Daemi:2003qd}. The latter emerges from a
6-dimensional abelian Higgs model coupled to gravity in a
Nielsen-Olesen vortex background with fermions. The effective
4-dimensional gauge field is a mixture of the original 6-dimensional
metric and the vector potential.

We determined the contribution of the extra dimensions to the
dispersion relations of the electromagnetic modes near the core of
the vortex, our world, and we found two types of contributions, Eq.
(\ref{dispersion}): a continuum one, associated with the radial
extra dimension  and a discrete one corresponding to a vortex
number. This behaves as $n^{-2}$ just as in the Casimir force for
compact non-commutative extra dimensions \cite{Nam:2000cv}. As a
result we get an effective Casimir energy, Eq. (\ref{eq:evortex}),
which differs with respect to the standard one by a multiplicative
factor rather than an additive term. This correction depends on both
parameters of the vortex scenario, namely the size of the core $a$
and the coupling constants length $\ell$. In the approximation
$l/\ell\ll 1$ the effective force, Eq. (\ref{fvortex}), contains
both an  attractive and a repulsive contributions with inverse
powers of the separation between plates $l^{-5}$ and $l^{-3}$,
respectively. Demanding agreement of this force with the experiment
to set bounds for the parameters of the effective QED does not work
due to the fact that the correction is multiplicative yielding a
different power in $l$ with respect to the standard case. This
limits the phenomenological implications of the effective QED here
considered.

 The appearance of the extra dimensional correction as a multiplicative factor
 that depends on the separation between plates seems to be a generic
 feature of non compact extra dimensions. Further studies in this
 direction for different models is in progress and will be reported
 elsewhere.

\section*{Acknowledgments}

\noindent We are grateful to Kei-ichi Maeda for useful discussions
and Jorge Alfaro and Kim Milton for enlightening observations. This
work was partially supported by Mexico's National Council of Science
and Technology (CONACyT), under grants CONACyT-40745-F,
CONACyT-NSF-E-120-0837 and SEP-2004-C01-47597. The work of O.P. was
supported by CONACyT Scholarship number 162767. He would also like
to thank the Young Collaborator Programme of Abdus Salam ICTP
Trieste, Italy, for supporting a visit  where part of this work was
developed.

\bibliography{plb05-058}
\bibliographystyle{toine}

\end{document}